\documentclass[preprint,12pt]{elsarticle}

\usepackage{graphicx}

\usepackage{amssymb}
\usepackage{amsthm}

\usepackage{bbm}
\usepackage{a4}
\usepackage[utf8]{inputenc}
\usepackage{enumerate}
\usepackage{epsf}
\usepackage{amsmath}
\usepackage{listings}

\newcommand{\newCSR}{Row-grouped CSR}
\newcommand{\NEWCSR}{RgCSR}

\journal{Parallel Computing}

\begin{document}

\begin{frontmatter}


\title{New \newCSR{} format for storing the sparse matrices on GPU with implementation in CUDA.} 
\ead{tomas.oberhuber@fjfi.cvut.cz}
\ead[url]{http://geraldine.fjfi.cvut.cz/~oberhuber}



\author[label1]{Tom\'{a}\v{s} Oberhuber}
\author[label2]{Atsushi Suzuki}
\author[label1]{Jan Vacata}
\address[label1]{Department of mathematics, Faculty of Nuclear Sciences and Physical Engineering, Czech Technical University in Prague, Trojanova 13, Praha 2, 120 00, Czech Republic}
\address[label2]{Laboratoire Jacques-Louis Lions, Universit\'e Pierre et Marie Curie, Bo\^ite courrier 187, 75252 Paris Cedex 05, France}

\author{}

\address{}

\begin{abstract}
In this article we present a new format for storing sparse matrices. 
The format is designed to perform well mainly on the GPU devices.
We present its implementation in CUDA.
The performance has been tested on $1,600$ different types of matrices and we compare our format with the Hybrid format proposed in \cite{BellGarland-2008}. 
We give detailed comparison of both formats and show their strong and weak parts.

\end{abstract}


\end{frontmatter}

\lstset{language=C,
        basicstyle=\footnotesize,
        numbers=left,
        frame=single,
        firstnumber=auto}


\section{Introduction}
\label{}
Graphics processing units (GPUs) are understood nowadays rather as high performance computational devices than only computer graphics accelerators. 
Their peak performance is beyond 1 TFLOPS in case of the single precision arithmetic. 
 Comparison of wide class of problems solved on the CPU and the GPU can be found in \cite{LeeKimChhuganiDeisherKimNguyenSatishSmelyanskiyChennupatyHammarlundSinghalDubey-2010}.
In this text, we concentrate on a kernel code of the numerical linear algebra.
 The dense matrix operations are successfully covered by BLAS implemented on GPU, e.g. CUBLAS \cite{NvidiaCublas-2010}.
Dense linear system solvers are presented in \cite{GaloppoGovindarajuHensonManocha-2005,VolkovDemel-2008}.
For the sparse matrices, solvers based on the Krylov subspaces methods are usually used.
This kind of iterative solvers spend most of the time by computing product of the matrix and vector.
This operation is denoted as SpMV (Sparse Matrix Vector multiplication).
Implementing iterative solvers like CG or GMRES on the GPU mainly requires having implemented SpMV (Sparse Matrix Vector multiplication), SAXPY (Scalar Alpha X Plus Y) and SDOT (Scalar DOT product) operations on GPU.
The rest of the solver remains the same as for the CPU implementation.
In this article we deal with the SpMV operation.
SAXPY and SDOT are rather simple operations and belong to BLAS level 1.

The GPU devices can profit from their great performance only in the case of arithmeticaly intesive algoritms.
Arithmetic intensity is a ratio of the number of arithmetic operations to memory accesses. 
SpMV operation $y=Ax$ for the sparse matrix $A$ requires one multiplication and one addition for each non-zero element. 
We need to read at least one non-zero element from the matrix $A$ and one element from the vector $x$.
The arithmetic intensity can not be higher than one.
It means that we are not bounded by the arithmetic performance of the GPU but by the memory bandwidth.

Formats for storing the sparse matrices often involves additional information which must be read.
Sometimes data need to be aligned for faster transfer which means adding articifial zero elements.
Both increase the amount of data to trasnfer and slow down the SpMV operation.
On the other hand, depending on the sparse matrix pattern, elements of the vector $x$ may be accessed repetitively.
Caching of the vector $x$ can improve the performance significiantly.
Efficient format for the sparse matrices should satisfy the following:

\begin{itemize}
\item store the data in continous blocks
\item store as less data as possible
\item reuse data of the vector $x$
\end{itemize}
This problem is relatively simple to solve when $A$ is well structured, for example if $A$ is multi-diagonal \cite{BellGarland-2008}.
For SpMV operation of general matrices, techniques for cache utilizations 
by block access \cite{Im-2000} and data compression for both index and 
value \cite{KourtisGoumasKoziris-2008} were introduced.
On the GPUs \cite{BaskaranBordwaker-2008} studied the SpMV operation for 
general matrix and \cite{ChoiSinghVuduc-2010} for sparse matrices 
appearing in the graph mining.
Hybrid format was proposed by \cite{BellGarland-2008} and was 
extended to a block version by \cite{MonakovAvetisyan-2009a}.

\subsection{Contributions}
We modify common CSR format to run efficiently on the GPU.
The format we obtain is simple as well as the kernel for the SpMV operation.
We have tested this format on 1600 sparse matrices from \cite{DavisHu-1994,BaiDayDemmelDongarra-1996}.
We present several statistics made from our experiments.
We compare the new format with the Hybrid format \cite{BellGarland-2008} and show strong and weak parts of both algorithms.
\subsection{Organization}
The article  is organized as follows. 
In the Section \ref{sect:cuda} we explain the necessary knowledge of CUDA device, which is a de facto standard of GPU computing device.
%
We establish what conditions should be fulfilled by the algorithm to gain maximum performance in SpMV operation on CUDA device.
In the Section \ref{sect:sparse-matrices-on-gpu} we show already existing formats and we study their advantages and disadvantages. 
In this section we also present the \newCSR{} format. 
The performance and comparison with the Hybrid format is subject to the Section \ref{sect:epxerimental-evaluation}. 
Here we show detail analysis of both formats from the performance point of view with both single and double precision.
\section{CUDA architecture}
\label{sect:cuda}
CUDA (Compute Unified Device Architecture) is an architecture designed by the Nvidia company to simplify the development of applications using GPU.
CUDA is restricted only to GPUs by Nvidia.
Geforce GTX280 is one of the first CUDA devices capable of computations in the double precision.
It is composed of 30 {\it multiprocessors} each having eight CUDA cores executing the single precision arithmetic and one processing unit for the double precision arithmetic.
Each multiprocessor is equipped with very fast 16 KB of {\it shared memory}.
It stores both data and instructions.
All the multiprocessors are connected to the {\it global memory},
which is understood as an SMP architecture.
%
Size of the global memory can be up to 1GB but it is only balanced with $20\%$ of peak performance of the double precision arithmetic.
There is a read only cache memory called a {\it texture cache}, which is bound to a part of the global memory when a code starts by the multiprocessors.

From the programmer's point of view, the most important computing entity is the {\it thread}.
One writes a code called {\it kernel} which is processed by many threads.
The multiprocessor can process 32 threads simultaneously.
Such group of threads is referred as {\it warp}.
Each thread of the warp must perform the same instruction at the same time.
The essential property of the CUDA threads is that they are very lightweight.
The multiprocessor is therefore capable to hold more than 32 threads and to switch between them efficiently.
This group of threads is called {\it block}.
The thread scheduler decides which threads are ready to be processed.
By this mechanism, the latencies of the global memory can be efficiently hidden.
There can be up to 512 threads in one block.
Threads can be explicitly synchronized by the programmer.
Blocks are grouped into {\it grids} and execution of blocks in the grid is distributed on multiprocessors. 

Fast access to global memory is essential for computation with CUDA device. 
%
%
This memory is well designed for sequential access but not for random access.
By nature of the hardware, the global memory is accessed by every aligned
128 bytes segment and is fed to threads in a half of the warp which is called
as {\it half-warp} (see \cite{CUDA-Prog-Guide-3-0-2010}).
The way how the the threads in the half-warp access to aligned 128 bytes segment is called as a {\it coalesced memory access}. 
%
For example, sequential access to array with 32 double precision variables
can be coalesced with halves.
When threads in a half-warp access to scattered address in 128$\times n$ bytes, 
128 bytes segments are accessed $n$ times, and as the result, memory access 
becomes $n$ times slower.
This memory access is called as non-coalesced.
Access to non-aligned 128 bytes segment is not coalesced either, because two times
access with 128 bytes are necessary.
Therefore, the programmer must design his code to fulfill the conditions of
coalesced memory access to obtain good performance.
With the coalesced memory accesses we can transfer more than 140 GB/s between the global memory and the multiprocessors. 
If one can find data reuse in the algorithm, the shared memory is better to be utilized with explicit code to copy data from the global memory into shared memory and to write back data.
Texture cache can improve non-coalesced memory access, although it is only valid for data reading.

The new matrix format, we present in this article, is based on reorganization of the CSR format such that the most of the data accesses can be coalesced. 

\section{Sparse matrix formats for GPUs}
\label{sect:sparse-matrices-on-gpu}
\subsection{Common CSR format}
\begin{figure}[]
\center{
\includegraphics[width=10cm]{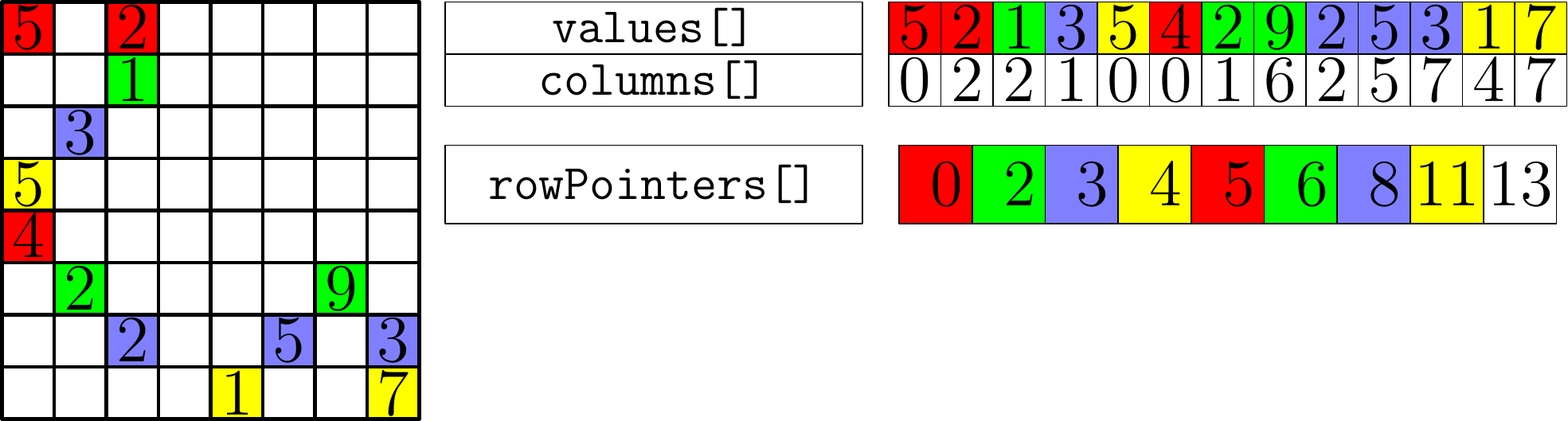}
}
\caption{CSR format}
\label{fig:csr}
\end{figure}

For common sequential systems the CSR (Compressed Sparse Rows) format is very popular.
It is because the matrix rows are easily accessible and it allows to write efficient code in memory usage for the SpMV.
The essence of the format is depicted on the Figure \ref{fig:csr}.
We store only the non-zero elements in a rowise order. 
Two arrays, {\tt values} and {\tt columns}, 
whose size is equal to the number of nonzero elements, 
store value and column index of the element in increasing order of the 
column index in each row.
%
An array {\tt rowPointers} keeps index of array where {\tt values} 
and {\tt columns} start to keep data in the row.
The {\tt rowPointers} has the same size of the matrix plus one, where
the last value equals the number of total non-zeros.
%

The code for multiplication of the sparse matrix by a vector looks like this:
%
\begin{lstlisting}
void spmvCSR( const int rowSize,
              const REAL* values,
              const int* columns,
              const int* rowPointers,
              const REAL* x,
              REAL* Ax )
{
   for( int i = 0; i < rowSize; i ++ )
   {
      Ax[ i ] = 0.0;
      int j;
      for( j = rowPointers[ i ]; j < rowPointers[ i + 1 ]; j ++ )
         Ax[ i ] += values[ j ] * x[ columns[ j ] ];
   }
}
\end{lstlisting}
There are two possibilities to parallelize this code.
Both of them are mentioned in \cite{BellGarland-2008}.
The first one is to use several threads per one row.
Each thread multiplies one non-zero element of the matrix $A$ with appropriate element of the vector $x$.
There are two main disadvantages of this approach.
We must implement relatively complicated parallel reduction.
Moreover, if there are only few non-zero elements on each row we do not have enough work for each thread of one warp.
The results presented in \cite{BellGarland-2008} show that this approach does not perform reasonably.
The authors refer it as vector CSR.
The second way is to map one thread per each row.
For large matrices there will be enough threads for efficient run on the GPU.
An advantage is that we do not need to change the code for the SpMV kernel.
On the other hand the way the threads read data from the arrays {\tt values} and {\tt columns} stored in the global memory does not fulfill the condition for the coalesced memory access.
It slows down this algorithms significantly.
In \cite{BellGarland-2008} it is referred as scalar CSR.
Both the scalar and the vector CSR are the slowest algorithms.
\subsection{Blocked CSR format}
\begin{figure}[]
\center{
\includegraphics[width=10cm]{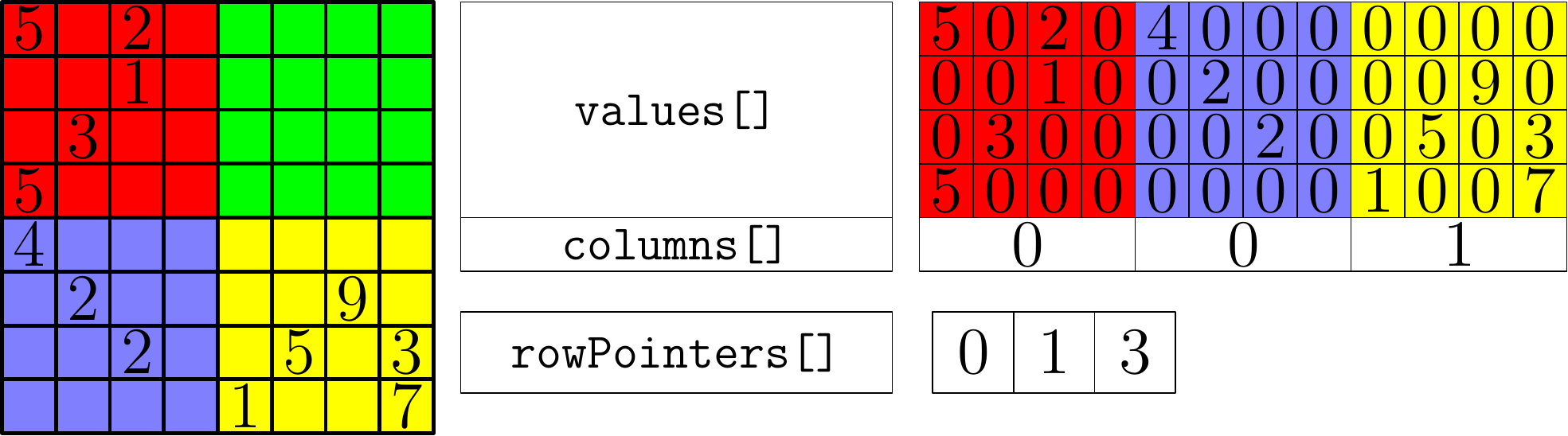}
}
\caption{Blocked CSR format}
\label{fig:bcsr}
\end{figure}

The authors of \cite{BuatoisCaumonLevy-2009} present Block CSR format developed for the vector-GPU architecture ATI and CTM (Close to Metal) \cite{AMD-2006}.
The authors decompose the matrix into 4x4 blocks as the Figure \ref{fig:bcsr} demonstrates.
The disadvantage of this format is that it decomposes into blocks the matrix itself and not the compressed sparse rows.
Because of this, many artificial zeros appear there.
As one can see on the Figure \ref{fig:bcsr}, there $48$ values but only $13$ of them are non-zero.
The efficiency of this storage is only $27\%$.
All these artificial zeros must be transfered from the global memory of the GPU which slows down the algorithm for SpMV operation.
It is also wasting of the GPU global memory.
Moreover the efficiency of this format decreases with larger block size.
For efficient use of the CUDA multiprocessors we would need block size equals $32$ for coalesced memory access.

\subsection{Hybrid ELLPACK and COO format}
\begin{figure}[]
\center{
\includegraphics[width=8cm]{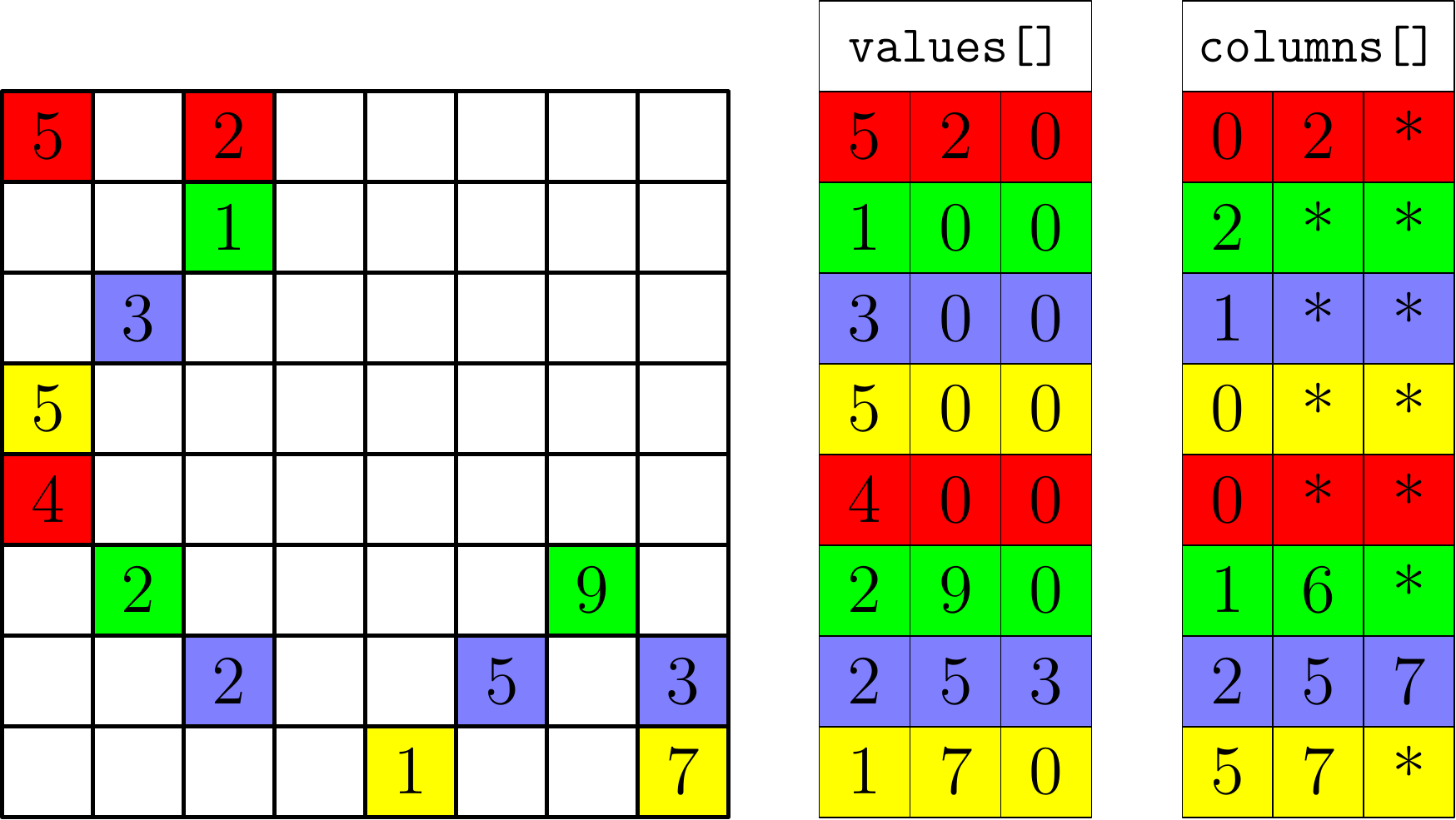}
}
\caption{ELLPACK format}
\label{fig:ellpack}
\end{figure}

\begin{figure}[]
\center{
\includegraphics[width=12cm]{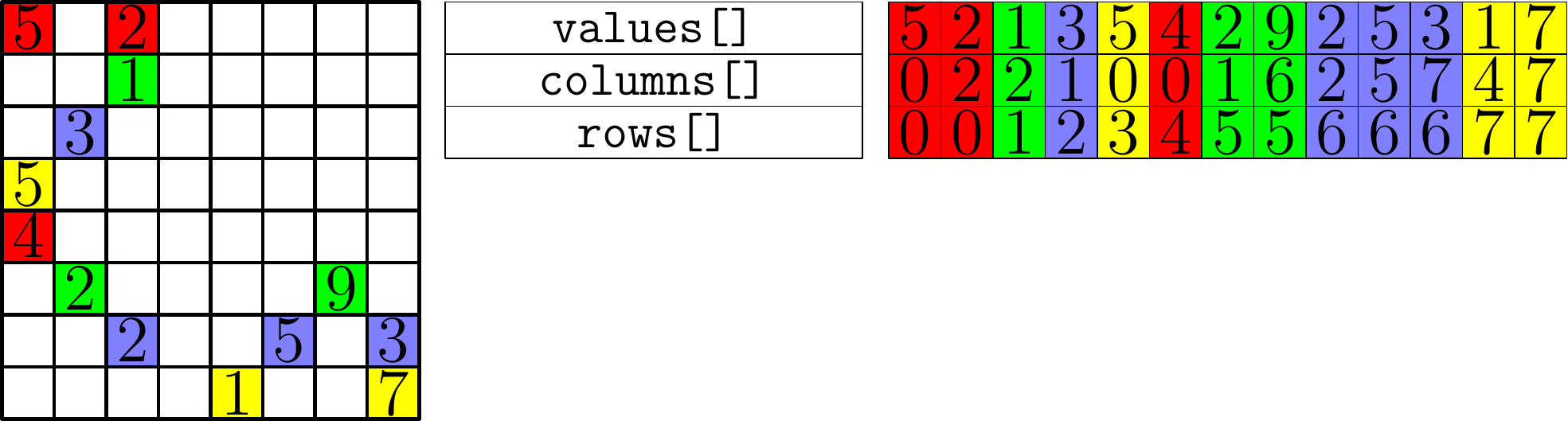}
}
\caption{COO format}
\label{fig:coo}
\end{figure}

Better format for the sparse matrices storage in the GPU is the Hybrid format introduced in \cite{BellGarland-2008}.
It is based on the combination of the ELLPACK and COO format.
For matrix with $N$ rows and with maximum $K$ non-zero elements per row, ELLPACK allocates $NK$ elements in the arrays {\tt values} and {\tt columns}.
Data storage is depicted on the Figure \ref{fig:ellpack}.
This format works well for matrices with approximately the same number of non-zero elements on each row.
Note that we do not need to store the {\tt rowPointers}.
Difficulties may appear when the number of non-zero elements differ significantly for each row.
Imagine that we have diagonal matrix which has one row full of non-zero elements.
In this case we have $2N-1$ non-zero elements but the ELLPACK format will store $N^2$ elements. 
Therefore the authors of \cite{BellGarland-2008} propose to combine the ELLPACK format with COO (coordinate) format.
The COO format is completely explicit format storing for each non-zero matrix element its row and its column - see Figure \ref{fig:coo}.
The Hybrid format allows to allocate less than $NK$ elements for the arrays {\tt values} and {\tt columns} of the ELLPACK format and those elements which do not fit into the allocated arrays are stored in the COO format.
The SpMV operation then consists of two steps, ELLPACK operation part and COO operation part.
Let's say that we allocate $NK_1$ elements for the ELLPACK format where $K_1 < K$.
If $K_1$ is only slightly smaller than $K$ then we may still have a lot of artificial zeros in the ELLPACK format.
If $K_1 \ll K$ then we may have a lot of non-zero elements stored in the COO format.
Since this format stores even the row coordinate for each non-zero element it requires more memory than CSR format.
Good choice of $K_1$ is essential for the Hybrid format.
\subsection{\newCSR{} format}
We present here new \newCSR{} format (we use RgCSR as its abbreviation). 
It is a simple modification \cite{Vacata-2008} of the common CSR format.
Independently, a very close format, sliced ELLPACK was published by \cite{MonakovLokhmotovAvetisyan-2010}. 
The problem is that the common CSR format do not fulfill the coalesced access 
to the array values and columns.
Let's consider the case that 8 threads proceed SpMV operation of 
common CSR format in Figure \ref{fig:csr} with mapping of one thread 
to one row of the matrix.
When each thread accesses to the first non-zero of each row, the positions 
in the arrays values and columns are 0, 2, 3, 4, 5, 6, 8 and 11. 
These elements are read at the same time.
We see that they are not accessed sequentially in the memory.
\par
\begin{figure}[]
\center{
\includegraphics[width=14cm]{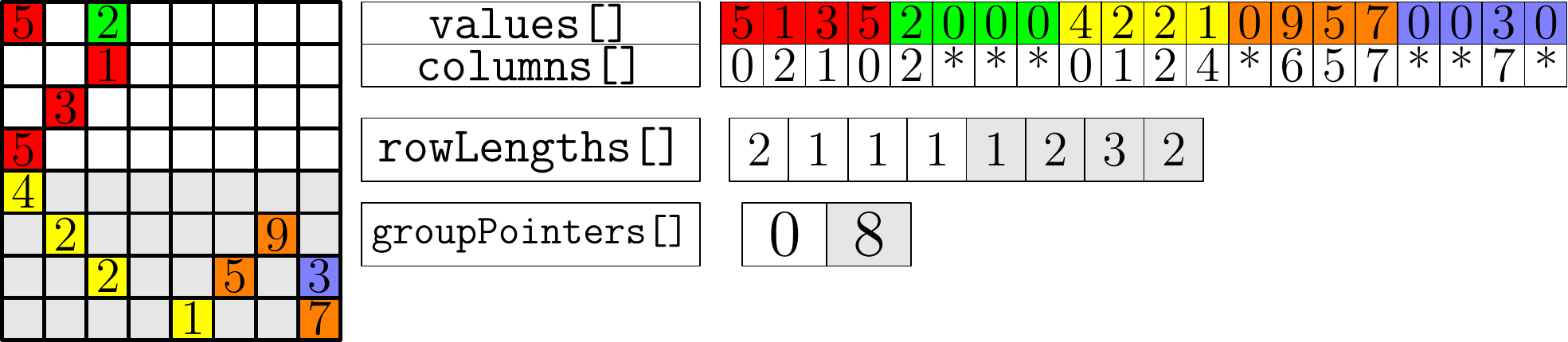}
}
\caption{\newCSR{} format}
\label{fig:RgCSR}
\end{figure}

The \newCSR{} format is based on storing these elements sequentially. We divide the matrix into groups of rows -- see the Figure \ref{fig:RgCSR}.
%
In this simple example we have two groups each having four rows.
In each group we store firstly the first non-zero elements in each row then the second non-zero elements in each row and so on.
If the number of the non-zero elements differs in some row of the group we add artificial zeros to have the same number of the elements to store in all rows of the group.
As the same way {\tt values} array keeps original non-zeros and artificial 
zeros, {\tt columns} array keeps index of column of each row and ghost index.
Instead of the {\tt rowPointers} array we store {\tt groupPointers} and {\tt rowLength} arrays. 
The first one keeps the offset of the group beginning in the {\tt values}/{\tt columns} arrays. 
The second one keeps the number of the non-zero elements in each row.
In implementation of SpMV with RgCSR format in Figure \ref{fig:RgCSR}, 
we can use 4 threads to one group.
\par
From the Figures \ref{fig:bcsr}, \ref{fig:ellpack} and \ref{fig:RgCSR} we see that the Blocked CSR format allocates $35$ artificial zeros, the ELLPACK format allocates $11$ and the \newCSR{} only $7$ of them.
An advantage of the \newCSR{} format over the ELLPACK format is that the number of allocated elements per one row may vary from one group to another.
\par
There is a difference between the \newCSR{} format and the
sliced ELLPACK format. 
Sliced ELLPACK does not store the number of non-zeros in each row, 
{\tt rowLengths[]}.
Maximum number of non-zeros in $j$-th strip, $K_{j}$ is calculated 
from indices of the first element of two strips,
{\tt groupPointers[j + 1]} $-$ {\tt groupPointers[j]}.
It gives uniform number of arithmetics in column direction for each strip. 
Sliced ELLPACK computes multiplications of zero-element and pseudo vector value to align arithmetic amount per row in the strip.
RgCSR format can skip such meaningless arithmetic by using explicit information of {\tt rowLengths[]}.
%
\par
The CUDA kernel for the \newCSR{} format reads as follows:

\begin{lstlisting}
__global__ void spmvRgCSR( const int rowSize,
                           const REAL* values,
                           const int* columns,
                           const int* groupPointers,
                           const int* rowLengths,
                           const REAL* x,
                           REAL* Ax )
{
  int row = blockIdx.x * blockDim.x + threadIdx.x;
  if( row >= rowSize ) return;

  int groupOffset = groupPointers[ blockIdx.x ];
  int ptr = groupOffset + threadIdx.x;

  // The last group may be smaller.
  int currentGroupSize = blockDim.x;
  if( ( blockIdx.x + 1 ) * blockDim. x > matrixSize )
          currentGroupSize = rowSize % blockDim.x;

  REAL product = 0.0;
  const int rowLength = rowLentghs[ row ];
  for( int i = 0; i < rowLength; i ++ )
  {
          product += values[ ptr ] * x[ columns[ ptr ] ];
          ptr += currentGroupSize;
  }
  Ax[ row ] = product;
}
\end{lstlisting}
%
Here, {\tt blockDim.x} takes size of block, which is set as the size of group. 
Integer variables {\tt blockIdx.x} and {\tt threadIdx.x} are index of block
whose takes between $0$ to $\lceil$ {\tt rowSize}$/$ {\tt blockDim.x} $\rceil{}-1$, 
and index of thread having $0$ to {\tt blockDim.x} $-1$, respectively.
The mathematical symobol $\lceil\,x\,\rceil$ shows the smallest integer grather or equal to $x$.
\par 
It is clear that the smaller group we have the less artificial zeros there are.
The smallest group size which can fulfill the condition of coalesced memory access  on the CUDA devices is $32$, i.e. the warp-size.
In practical computing we usually choose larger groups size.
\par
We now show estimation of the peak performance of the CUDA kernel for the \NEWCSR{}
format.
We assume the maximum memory performance as $m$ GB/sec.
For each non-zero element we need to perform one
multiplication and one addition. 
It means that the number of floating point
operations per one SpMV operation is twice the number of the non-zero elements
in the matrix. 
To process arithmetic for one non-zero element we must read one integer from
the {\tt columns} array and two single or double precision 
floating point numbers, where
one comes from the array {\tt values} and one from the vector $x$. 
For simplicity in estimation of upper bound of the performance, 
we omit the other arrays. 
%
Since 32 bit integer is used in CUDA GPU, 
data access on each step takes 12 bytes in the
single precision arithmetic and 20 bytes in the double precision arithmetic. 
We will attain the maximum performance as $m/12$ GFLOPS for single precision
and $m/20$ GFLOPS for double precision, respectively.
We note that access to the array values is coalesced by the design of 
the format, but the access to the vector $x$ data is not coalesced in general.
Therefore, real performance is deteriorated by non-coalesced access to
the $x$ data.
For remedy of this problem, we can utilize cache memory for reading vector $x$.
This is done by binding the vector $x$ to a texture in CUDA device.
In the ideal case of perfect data-reuse of the vector $x$, almost all 
data accesses of $x$ are cached and we can omit reading the vector $x$ from our estimation.
This leads to 8 bytes in the single precision arithmetic and
12 bytes in the double precision arithmetic.
The maximum performance will be
$m/8$ GFLOPS for single precision
and $m/12$ GFLOPS for double precision, respectively.
Table \ref{table:estimation-peak-GPU} shows estimation of the peak performance of the GTX280 card
with 141 GB/s bandwidth.
\begin{table}[b]
\begin{center}
  \begin{tabular}{|c|cc|}
    \hline
    texture cache for vector $x$ & single & double \\ 
    \hline    \hline
    without  & 23.5 & 14.1 \\
    with  & 35.25  & 23.5\\
    \hline
  \end{tabular}\caption{Estimation of peak performances of the \newCSR{}
  format on GTX280 with 141GB/sec memory bandwidth.}
\label{table:estimation-peak-GPU}
\end{center}
  \end{table}
\section{Experimental evaluation}
\label{sect:epxerimental-evaluation}
\subsection{Setting of experiments}
The experiments were performed on the PC equipped with 
Intel Core2 Quad CPU Q6700 running at 2.66GHz with 4MB L2 Cache,
8 GB DDR3-1333 SDRAM, and Nvidia GTX 280 card.
While Nvidia GTX 280 has the peak memory bandwidth 141 GB/s,
the DDR3-1333 module has 10.667 GB/s. 
We used CUDA toolkit ver.3.1 to implement our \newCSR{} format 
and an implementation of the Hybrid format from CUSP library \cite{Nvidia-2010}.
\par
Doing the same estimation of the peak performance of the CSR format on the CPU as we did for the \newCSR{} format on
GTX280, we get 0.89 GFLOPS for the single precision and 0.53 GFLOPS 
for the double precision without cache memory.
Since the cache memory effect of the CPU is much more complicated than on the GPU,
it is difficult to estimate the peak performance of the CSR format 
on CPU with cache memory %
(\cite{NishtalaVuducDemmelYelick-2007} provides an estimation 
for the CSR format with block access).
However, we can see there is big advantage of GPU to CPU.
\par
We have tested the common CSR, the Hybrid and the \newCSR{} formats on a set of $1,596$ square matrices collected from two matrix markets \cite{DavisHu-1994,BaiDayDemmelDongarra-1996}.
The statistics were computed in three ways - on the complete set of matrices, on small matrices with size smaller than $10,000$ and on large matrices with size larger or equal $10,000$.
The Table \ref{matrix-sets-properties} shows minimum, maximum and average size, non-zero elements and ratio of non-zeros to number of the whole elements of the matrix in each set.
\begin{center}
\begin{table}
\begin{tabular}{|l|r|r|r|}
\hline
                      & Complete set        & Small matrices & Large matrices      \\
\hline \hline
Number of matrices    & $1598$              & $1061$      & $537$                  \\  
\hline
Min. size             & $5$                 & $5$         & $10,000$               \\
Max. size             & $2,063,494$         & $9,941$     & $2,063,494$            \\ 
Average size          & $41,258$            & $3,253$     & $116,059$              \\
\hline 
Min. non-zero els.    & $15$                & $15$        & $6,639$                \\
Max. non-zero els.    & $52,672,325$        & $3,279,690$ & $52,672,325$           \\
Average non-zero els. & $947,367$           & $64,899$    & $2,684,263$            \\
\hline
Min. non-zero ratio   & $3\cdot 10^{-6}\%$  & $0.01\%$    & $3\cdot 10^{-6}\%$     \\
Max. non-zero ratio   & $100\%$             & $100\%$     & $2.10\%$                \\
Average non-zero ratio& $1.20\%$             & $1.76\%$    & $0.09\%$               \\
\hline
\end{tabular} 
\caption{Properties of the matrix sets.}
\label{matrix-sets-properties}
\end{table}
\end{center}
\begin{center}
\begin{table}
\begin{tabular}{|l||r|r||r|r||r|r||}
\hline
                      &\multicolumn{2}{|c||}{Complete set}
                      &\multicolumn{2}{|c||}{Small matrices}
                      &\multicolumn{2}{|c||}{Large matrices}      \\
\hline \hline
                     & Single    & Double    & Single    & Double  & Single  & Double \\
\hline
CSR min.             & $0.16$    &  $0.1$    & $0.1$     & $0.1$   & $0.19$  & $0.19$ \\
CSR max.             & $1.4$     &  $1.3$    & $1.4$     & $1.3$   & $1.2$   & $1.2$  \\ 
CSR average          & $0.88$    &  $0.83$   & $0.84$    & $0.83$  & $0.82$  & $0.82$ \\
\hline                                                                                
Hybrid min.          & $0.0009$  &  $0.0008$ & $0.0009$  & $0.0008$& $0.24$  & $0.19$ \\
Hybrid max.          & $16.0$    &  $11.0$   & $8.9$     & $6.1$   & $16.0$  & $11.0$ \\
Hybrid average.      & $2.56$    &  $1.57$   & $1.07$    & $0.72$  & $5.48$  & $3.2$  \\
\hline                                                                                
Speed-up min.        & $0.00021$ &  $0.003$  & $0.00021$ & $0.003$ & $0.63$  & $0.48$ \\
Speed-up max.        & $36$      &  $10$     & $7.2$     & $4.9$   & $36$    & $11$   \\
Speed-up average     & $2.54$    &  $1.76$   & $0.97$   & $0.69$  & $5.59$  & $3.87$ \\
\hline
\end{tabular} 
\caption{Performance of the common CSR format and the Hybrid format in GFLOPS. Speed-up of the Hybrid format is measured against the common CSR format.}
\label{performance-csr-hybrid}
\end{table}
\end{center}
\subsection{Peformance of Hybrid format}
We first show performance of the common CSR format and the Hybrid format on GTX 280 card in Table \ref{performance-csr-hybrid}.
We see that the performance of the common CSR format does not depend much neither on the size of the matrix nor on the precision of the arithmetic.
The maximum performance of the Hybrid format is 16 GFLOPS in the single precision resp. 11 GFLOPS in the double precision.
It is $68\%$ resp. $78\%$ of the estimated performance of the GTX 280 card.
We also see that for the small matrices the average speed-up is 0.97 resp. 0.69.
So, in general it does not make sense to use the Hybrid format for the small matrices even though it can be $6$ or $7$ times faster in some special cases.
The situation is much better with the large matrices where the average speed-up is $5.59$ and it is $0.63$ in the worst case.
\subsection{Performance of \NEWCSR{} format}
The Table \ref{performance-RgCSR-csr} shows performance and the speed-up of the \newCSR{} on GTX280 with respect to various group sizes 32 to 256.
It also shows the filling of the \NEWCSR{} format with artificial zeros.
Let us start by commenting the filling.
$100\%$ filling means that there is the same amount of the artificial zeros as the non-zero elements.
The best filling is $0\%$.
It is attained when matrix has the same number of non-zero elements in every group of rows whose size equals to group size.
We do not show this in the table.
The best average value is $105\%$.
It means that in average we must store twice as much data as the 
common CSR format. 
%
In the worst case, we store $85$ times more data. 
We see this as major weakness of the \NEWCSR{} format. 
We can see that the rate of filling almost does not depend on the matrix size,
because filling is done as local operation with group size, which is much smaller than the matrix size.
\par
Now let us to see the effect of the group size on the performance.
We can see that up to the
group size $128$ the performance grows and it drops a little for 
the group size $256$.
%
Memory access with group size 32 satisfies coalesced access.
However there is another factor for faster memory access and
more than 32 threads (one warp) are necessary to hide memory latencies well.
How well the threads access the global memory can be measured by the warp 
{\it occupancy} (see \cite{CUDA-Prog-Guide-3-0-2010}).
With occupancy $1.0$ multiprocessors run warps without delay caused by
memory access.
In this experiment, occupancy is $0.25$ with 32 threads, $0.5$ with 64 threads,
and $1.0$ with 128 and 256 threads.
%
%
Due to increasing of number of filling, ratio of
effective memory in the coalesced is decreasing, so we have an optimal
size on the group.
%
%
%
%
\par
 In average, the difference in the performance is not so significant but for the maximum performance it is, especially for the single precision.
Here it grows from $20.6$ GFLOPS to $32.8$ GFLOPS. 
On the comparison of RgCSR on GPU to common CSR on CPU, RgCSR on GPU is $4.3$ times resp. $3.4$ times faster in average, but $100$ times slower in the worst case. 
  When we restrict ourselves only to small matrices these numbers decrease to $2.2$ resp. $1.87$.
It means that the \NEWCSR{} can be reasonably used even for small matrices for which the Hybrid format is not profitable choice.
For the large matrices the \NEWCSR{} also offers good performance, e.g., speed-up is 8.64 resp. 7.94, whlie 5.5 resp. 3.2 of Hybrid format. 
 
 We would like to note that \NEWCSR{} format on CPU has possibility to perform better than the common CSR.
In average it gives only $55\%$ performance of the CSR format but in some cases \NEWCSR{} format can better use the cache and it can be up to $4$ times faster than CSR.
However, we will not study the \NEWCSR{} on the CPU more in this text.

Let us also comment the effect of caching the vector $x$ by binding it to the texture memory.
Without caching the best performance in the single precision was only $18.03$ GFLOPS.
Turning the caching on increased this number $1.81$ times to $32.84$.
In average, the difference was up to $30\%$.
With the double precision the best performance grows $1.3$ times from $14.46$ GFLOPS to $18.82$ GFLOPS and in average the difference is $37\%$.

Another important information is how far from the peak performance we are.
If we take the best case, i.e. $32.8$ resp. $18.82$ GFLOPS for the single resp. the double precision, it makes $93\%$ resp. $80\%$ of the peak performance. 
If we omit the caching of the vector $x$ it is $53\%$ resp. $71\%$. 
\begin{center}
\begin{table}
\begin{tabular}{|l||r|r||r|r||r|r||}
\hline
                      &\multicolumn{2}{|c||}{Complete set}
                      &\multicolumn{2}{|c||}{Small matrices}
                      &\multicolumn{2}{|c||}{Large matrices}                                                                 \\
\hline 
                     & Single    & Double    & Single    & Double  & Single  & Double                                        \\
\hline \hline
                     &\multicolumn{6}{c||}{Group size 32}                                                                    \\
\hline
Artif. zeros max.    &\multicolumn{2}{|c||}{$1032\%$}  &\multicolumn{2}{|c||}{$1032\%$}    &\multicolumn{2}{|c||}{$870\%$}   \\
Artif. zeros average &\multicolumn{2}{|c||}{$105\%$}   &\multicolumn{2}{|c||}{$107\%$}     &\multicolumn{2}{|c||}{$102\%$}   \\
\hline
\NEWCSR{} min.             & $0.01$     &  $0.01$           & $0.01$      & $0.01$              & $0.02$      & $0.02$            \\
\NEWCSR{} max.             & $20.6$     &  $16.33$          & $11.2$      & $9.23$              & $20.6$      & $16.33$           \\ 
\NEWCSR{} average          & $3.19$     &  $2.72$           & $1.7$       & $1.5$               & $6.09$      & $5.14$            \\
\hline                                                                                
Speed-up min.         & $0.01$     &  $0.01$           & $0.01$      & $0.01$              & $0.02$      & $0.02$            \\
Speed-up max.         & $18.75$    &  $26.2$           & $9.1$       & $9.1$               & $18.75$     & $26.2$            \\
Speed-up average      & $3.18$     &  $2.72$           & $1.7$       & $1.5$               & $6.15$      & $6.27$            \\
\hline \hline
                      &\multicolumn{6}{c||}{Group size 64}                                                                   \\
\hline
Artif. zeros max.     &\multicolumn{2}{|c||}{$2098\%$} &\multicolumn{2}{|c||}{$2098\%$}    &\multicolumn{2}{|c||}{$1430\%$}  \\
Artif. zeros average  &\multicolumn{2}{|c||}{$144\%$}  &\multicolumn{2}{|c||}{$137\% $}    &\multicolumn{2}{|c||}{$157\%$}   \\
\hline
\NEWCSR{} min.             & $0.01$     &  $0.01$            & $0.01$     & $0.01$              & $0.02$      & $0.02$            \\
\NEWCSR{} max.             & $28$       &  $18.64$           & $19$       & $14.9$              & $28.05$     & $18.46$           \\ 
\NEWCSR{} average          & $4.18$     &  $3.35$            & $2.2$      & $1.9$               & $8.01$      & $6.19$            \\
\hline                                                                                
Speed-up min.         & $0.01$     &  $0.01$            & $0.01$     & $0.01$              & $0.02$      & $0.02$            \\
Speed-up max.         & $25.89$    &  $27.13$           & $15.4$     & $14.6$              & $25.9$      & $27.13$           \\
Speed-up average      & $4.14$     &  $3.81$            & $2.13$     & $1.9$               & $8.07$      & $7.55$            \\
\hline \hline
                      &\multicolumn{6}{c||}{Group size 128}                                                                  \\
\hline
Artif. zeros max.     &\multicolumn{2}{|c||}{$4230\%$}  &\multicolumn{2}{|c||}{$4230\%$}   &\multicolumn{2}{|c||}{$2476\%$}  \\
Artif. zeros average  &\multicolumn{2}{|c||}{$206\%$}   &\multicolumn{2}{|c||}{$183\%$}    &\multicolumn{2}{|c||}{$250\%$}   \\
\hline
\NEWCSR{} min.             & $0.01$     &  $0.01$            & $0.01$      & $0.01$             & $0.02$       & $0.02$            \\
\NEWCSR{} max.             & $32.8$     &  $18.82$           & $19.9$      & $14.15$            & $32.84$      & $18.82$           \\ 
\NEWCSR{} average          & $4.38$     &  $3.43$            & $2.2$       & $1.87$             & $8.58$       & $6.52$            \\
\hline                                                                                
Speed-up min.         & $0.01$     &  $0.01$            & $0.01$      & $0.01$             & $0.02$       & $0.02$            \\
Speed-up max.         & $26.98$    &  $26.22$           & $16.1$      & $13.94$            & $26.98$      & $26.22$           \\
Speed-up average      & $4.34$     &  $3.91$            & $2.14$      & $1.87$             & $8.64$       & $7.94$            \\
\hline \hline
                      &\multicolumn{6}{c||}{Group size 256}                                                                  \\
\hline
Artif. zeros max.     &\multicolumn{2}{|c||}{$8494\%$}  &\multicolumn{2}{|c||}{$8494\% $} &\multicolumn{2}{|c||}{$4684\%$}   \\
Artif. zeros average  &\multicolumn{2}{|c||}{$304\%$}   &\multicolumn{2}{|c||}{$255\%$}   &\multicolumn{2}{|c||}{$400\%$}    \\
\hline
\NEWCSR{} min.             & $0.01$     &  $0.01$            & $0.01$      & $0.01$            & $0.02$      & $0.02$             \\
\NEWCSR{} max.             & $31.6$     &  $18.3$            & $21.3$      & $13.9$            & $31.57$     & $18.3$             \\ 
\NEWCSR{} average          & $4.37$     &  $3.37$            & $2.24$      & $1.83$            & $8.55$      & $6.38$             \\
\hline                                                                                
Speed-up min.         & $0.01$     &  $0.01$            & $0.01$      & $0.01$            & $0.02$      & $0.02$             \\
Speed-up max.         & $26.1$     &  $25.46$           & $17.2$      & $13.7$            & $26.13$     & $25.46$            \\
Speed-up average      & $4.33$     &  $3.37$            & $2.15$      & $1.83$            & $8.58$      & $7.78$             \\
\hline
\end{tabular} 
\caption{Performance of the \NEWCSR{} format. Speed-up of the \NEWCSR{} format is measured against the common CSR format running on the CPU.}
\label{performance-RgCSR-csr}
\end{table}
\end{center}
\subsection{Comparison of \NEWCSR{} format and Hybrid format}
%
We show comparison of \NEWCSR{} format to others. Here we fixed group size as 128,
which attains the best results in Table 4.
%
\subsubsection{Outlines from the set of matrices}
\begin{center}
\begin{table}
\begin{tabular}{|l||r|r||r|r||r|r||}
\hline
                      &\multicolumn{2}{|c||}{Complete set}
                      &\multicolumn{2}{|c||}{Small matrices}
                      &\multicolumn{2}{|c||}{Large matrices}                                                                 \\
\hline 
                           & Single    & Double    & Single    & Double  & Single  & Double                                        \\
\hline                     
HYB faster than CSR        & 48.17\%   & 42.63\%   &  22.32\%  & 16.71\% & 98.70\% & 93.64\%                                       \\
\NEWCSR{} faster than CSR       & 56.68\%   & 55.79\%   &  46.34\%  & 45.20\% & 76.70\% & 76.64\%                                       \\
\NEWCSR{} faster than HYB       & 77.14\%   & 80.67\%   &  84.43\%  & 85.40\% & 62.57\% & 71.40\%                                       \\
Average. \NEWCSR{}/HYB          &  2.55\    &  3.21\    &   3.21\   &  3.74\  &  1.24\  &  2.18\                                          \\
\hline
\end{tabular} 
\caption{Comparison of the CSR, \NEWCSR{} and the Hybrid format. The first three rows say for how many matrices is one format faster than the others. It also shows average speed-up of the \NEWCSR{} format against the Hybrid format.}
\label{RgCSR-vs-hyb}
\end{table}
\end{center}

\begin{figure}[]
\center{
\includegraphics[angle=-90,width=12cm]{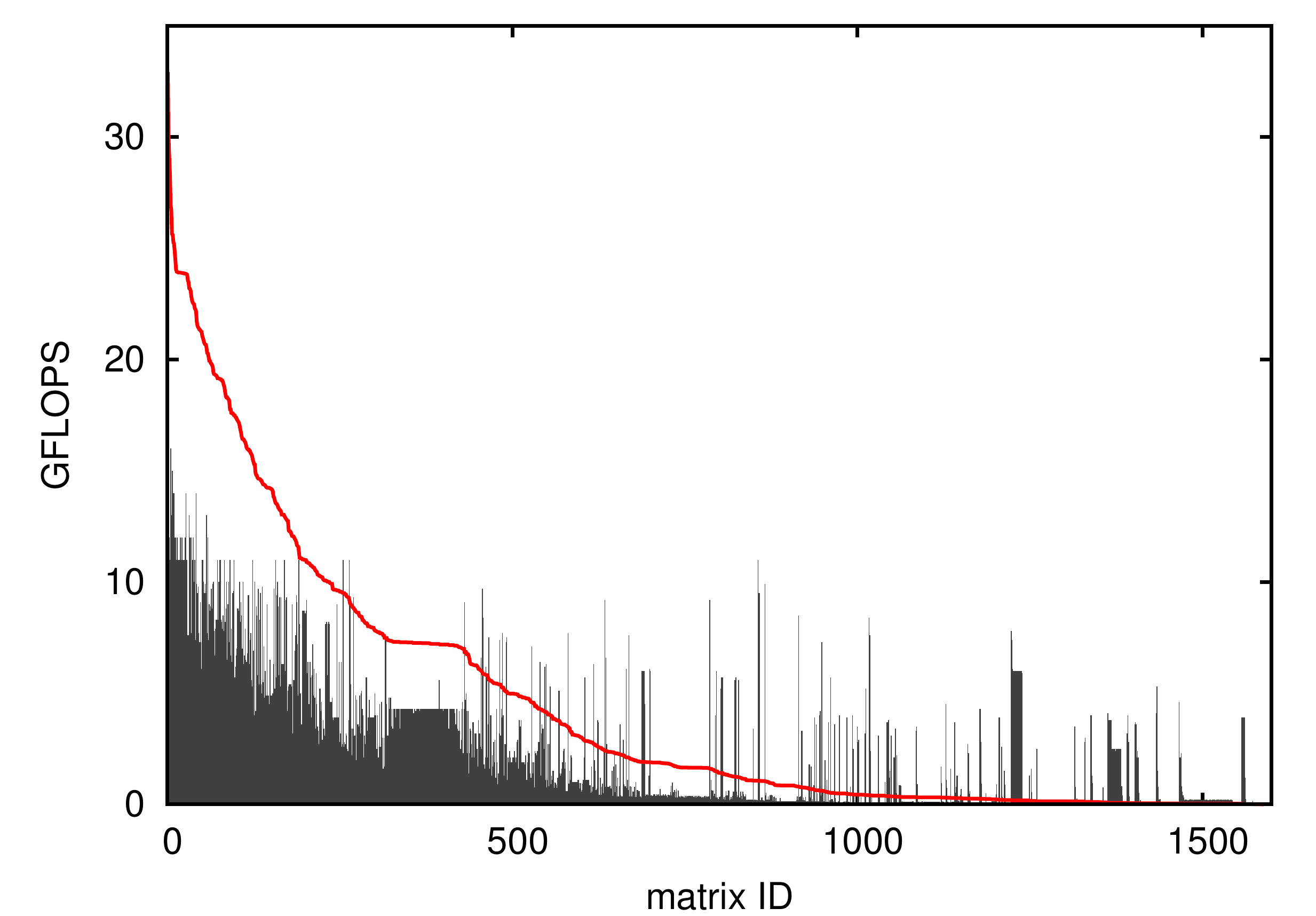}
\includegraphics[angle=-90,width=12cm]{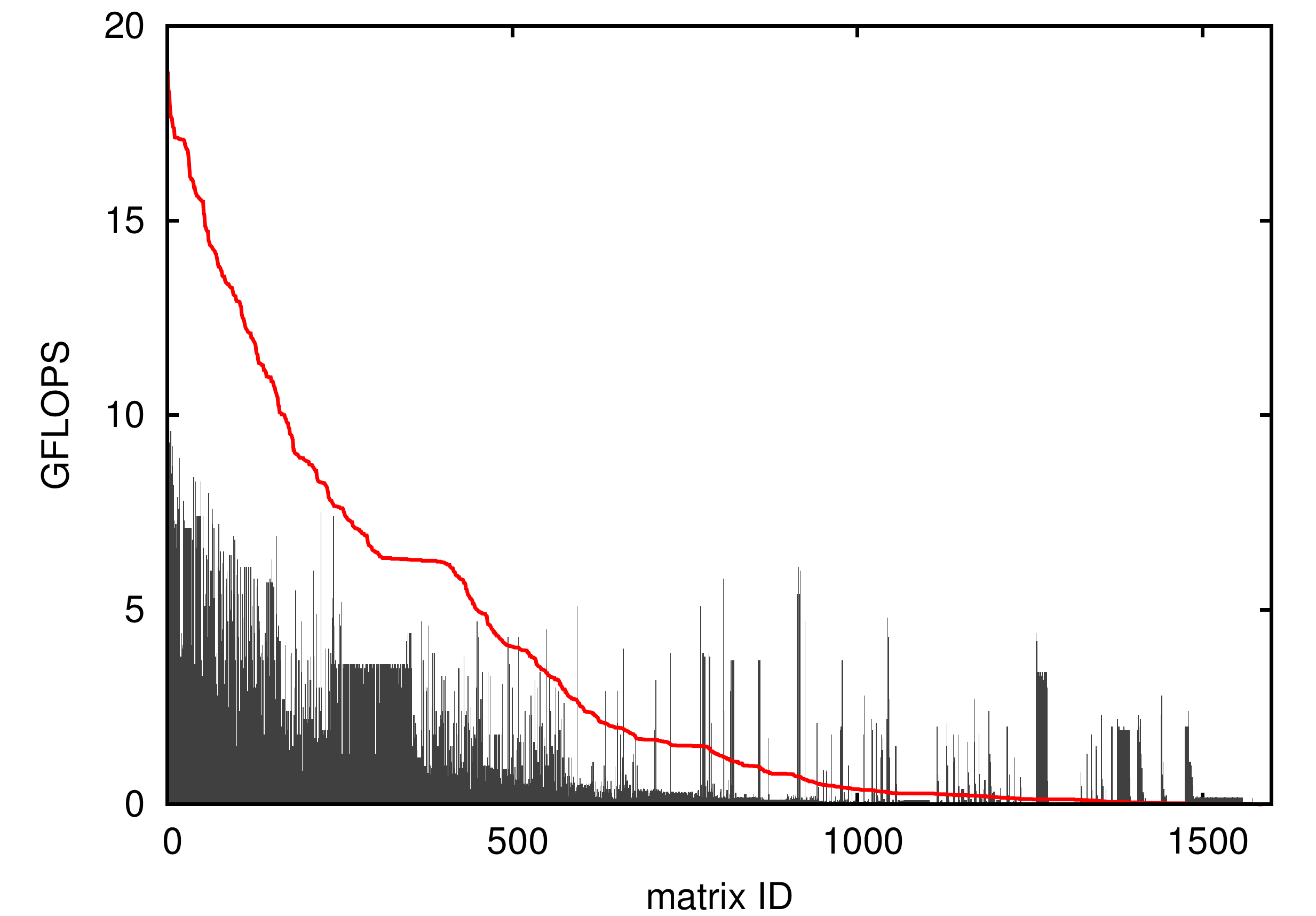}
}
\caption{Performance of both the \NEWCSR{} (red line) and the Hybrid (black bars) formats in the single (top) and the double (bottom) precision on the complete set of $1,596$ matrices.
The $x$-axis is the matrix ID and the $y$-axis is the performance in GFLOPS.
The matrices are sorted in the descending order by the performance of the \NEWCSR{} format.}
\label{fig:RgCSR-vs-hybrid}
\end{figure}

\begin{figure}[]
\center{
\includegraphics[angle=-90,width=12cm]{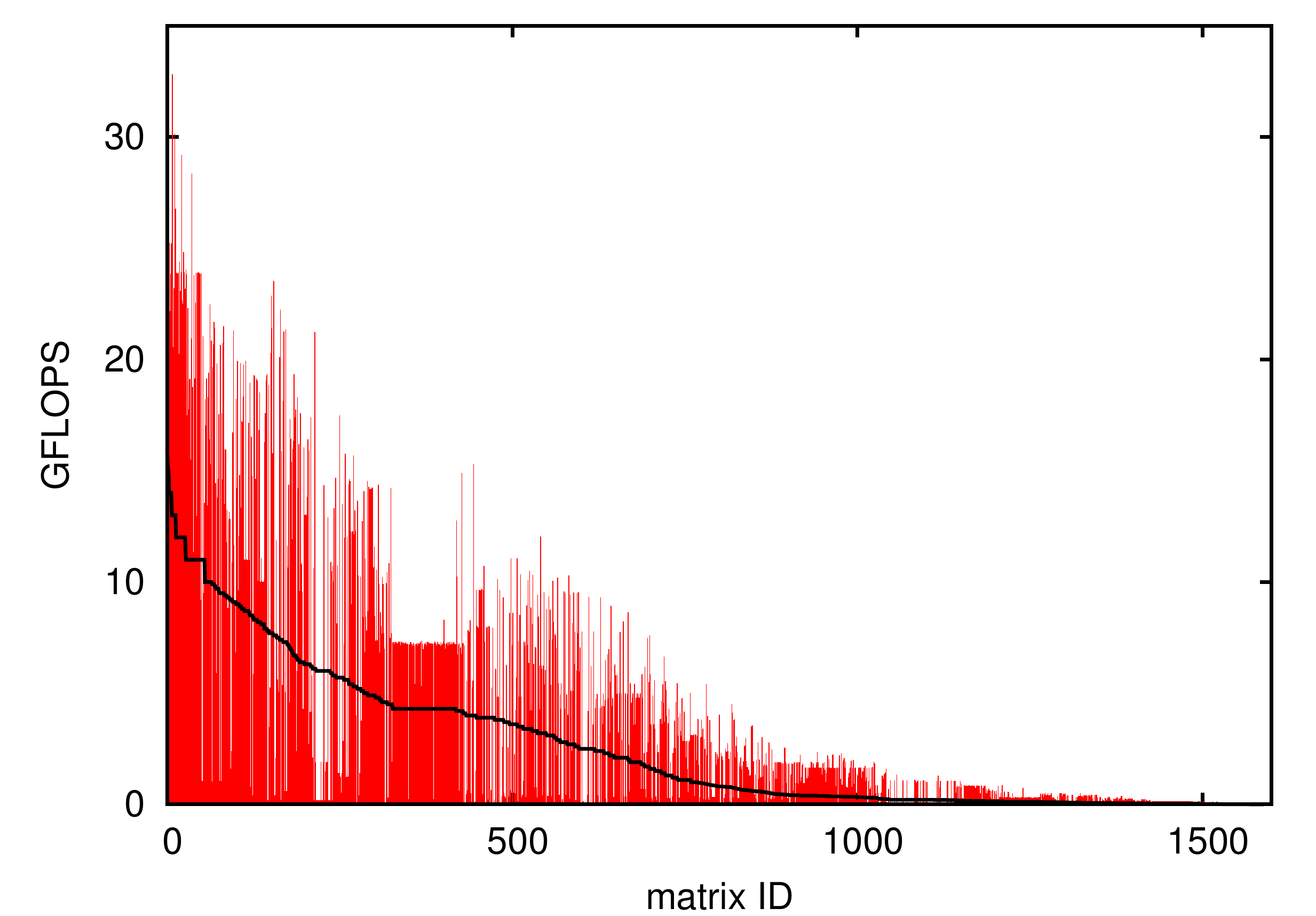}
\includegraphics[angle=-90,width=12cm]{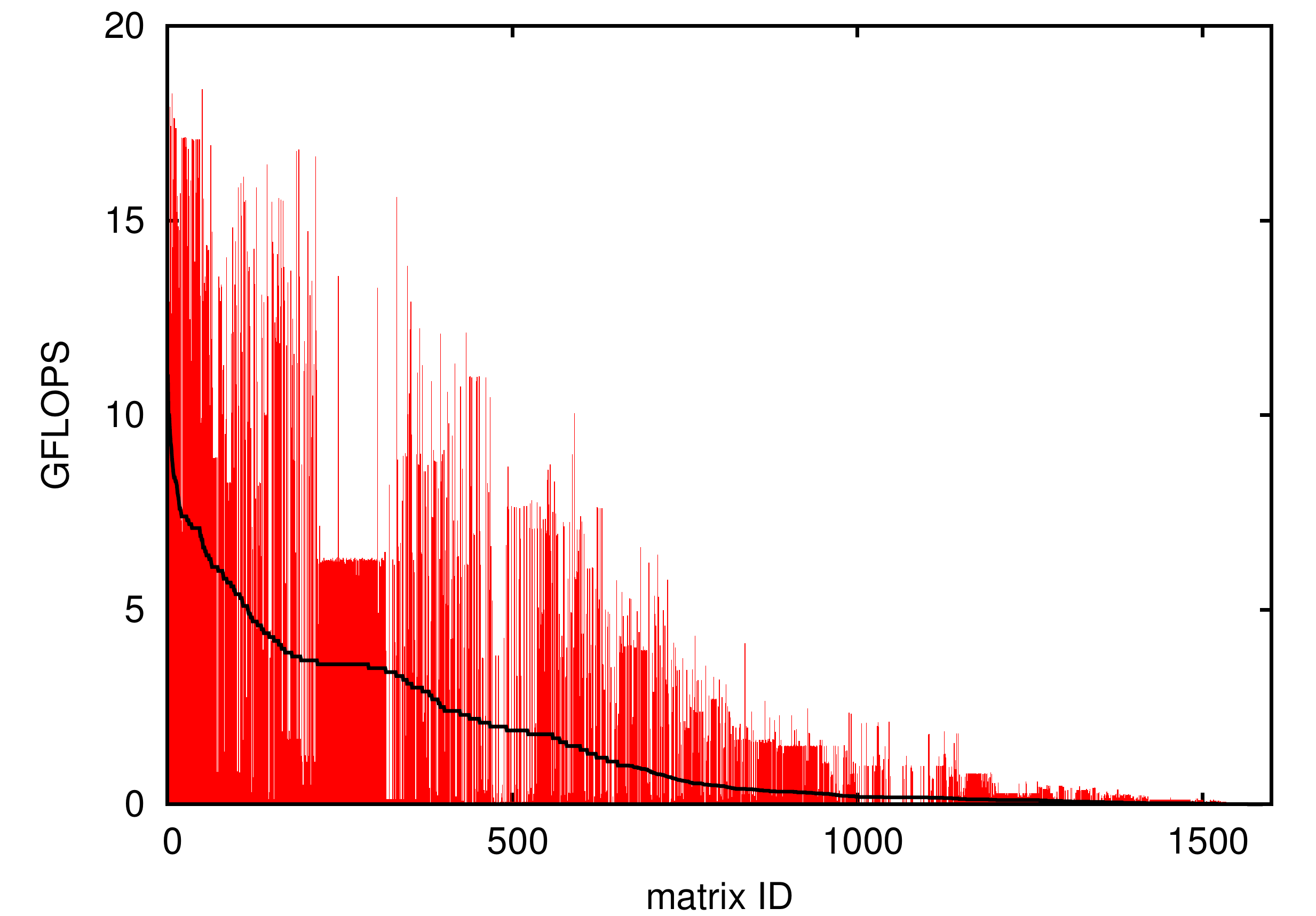}
}
\caption{``Inverse'' figure to the Figure \ref{fig:RgCSR-vs-hybrid-2}.
Performance of the \NEWCSR{} (black line) and the Hybrid (red bars) formats in the single (top) and the double (bottom) precision on the complete set of $1,596$ matrices.
The $x$-axis is the matrix ID and the $y$-axis is the performance in GFLOPS.
The matrices are sorted in the descending order by the performance of the Hybrid format.}
\label{fig:RgCSR-vs-hybrid-2}
\end{figure}

The Table \ref{RgCSR-vs-hyb} shows the comparison of the best \NEWCSR{} setting and the Hybrid format.
It shows in how many cases the Hybrid format is faster than the CSR format and the same for the \NEWCSR{} vs. CSR format and the \NEWCSR{} vs. the Hybrid format.
It also shows the average speed-up of the \NEWCSR{} format related to the Hybrid format.
Here we can again see, that the \NEWCSR{} format outperforms the Hybrid format well with the small matrices.
With the large ones the \NEWCSR{} format is $1.2$ resp. $2.18$ times faster than the Hybrid format.
\par
Figures \ref{fig:RgCSR-vs-hybrid} and \ref{fig:RgCSR-vs-hybrid-2} show 
detailed performance comparison of both RgCSR and the Hybrid formats with all 
1,596 matrices by graphs whose $x$-axis is based on sorted matrix-ID according
to either performance. 
In the single precision arithmetics there are only few matrices were \NEWCSR{} achieves more than 30 GFLOPS (to be more specific there are only 3 of them), there are 30 matrices where \NEWCSR{} gets over 20 GFLOPS and 200 matrices with the performance over 10 GFLOPS.
There are only 60 matrices where the Hybrid format gets over 10 GFLOPS.
Similar results can be observed even with the double precision -- there are $170$ matrices for which the \NEWCSR{} performs better than $10$ GFLOPS and only $4$ matrices for which the Hybrid format gets over $10$ GFLOPS.
We observe the following tendency in double precision from the graph
of Figure \ref{fig:RgCSR-vs-hybrid}, bottom.
Matrices whose computation speed with \NEWCSR{} is grater than $4$ GFLOPS
allow faster computation than the Hybrid format.
On the other hand, in case of matrices which obtain only less than $1$ GFLOPS
by \NEWCSR{}, the Hybrid format has possibility to perform faster than \NEWCSR{}.
%
%
\subsubsection{Detailed comparison with specific matrices}
\begin{figure}[]
\center{
\includegraphics[width=7cm]{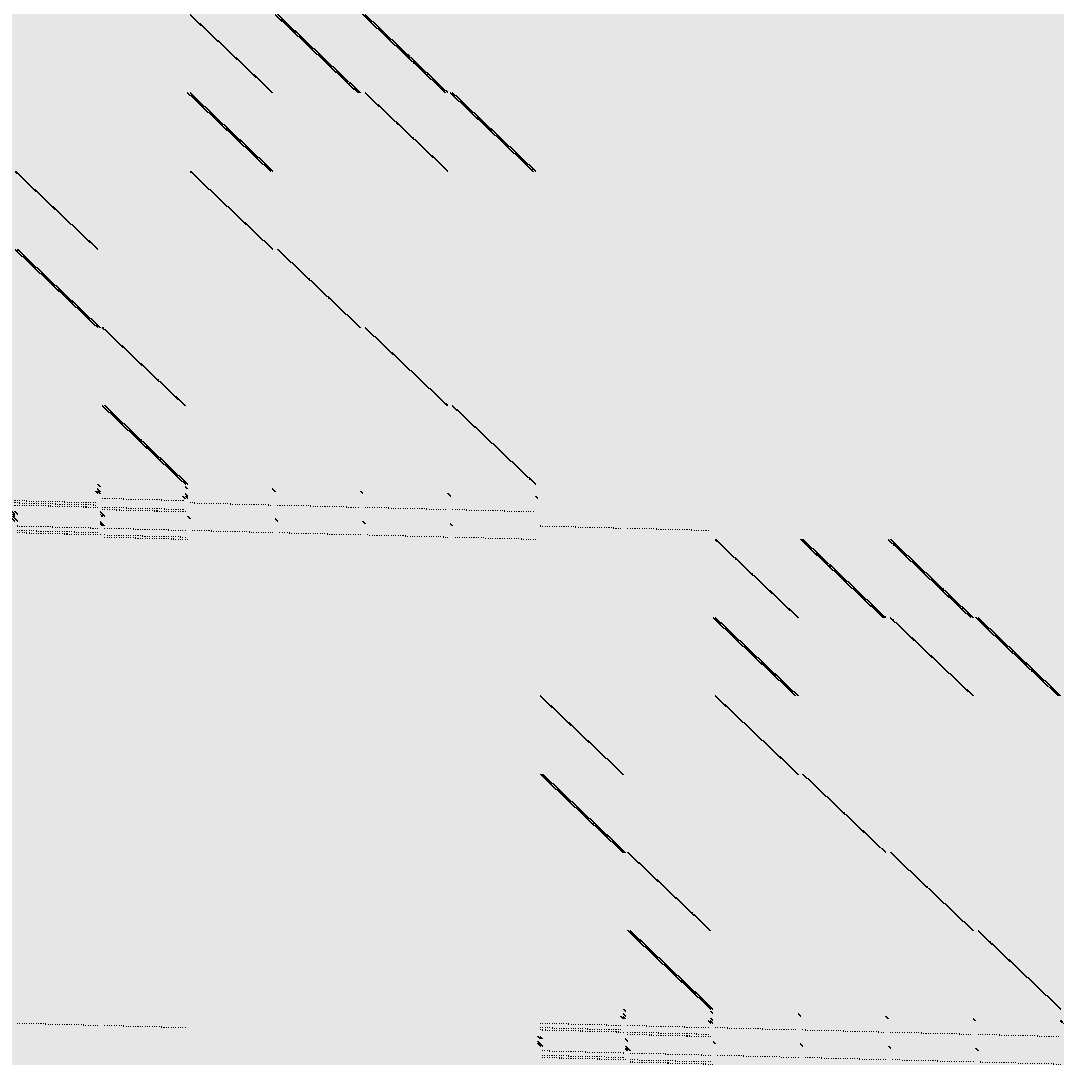}
\includegraphics[width=7cm]{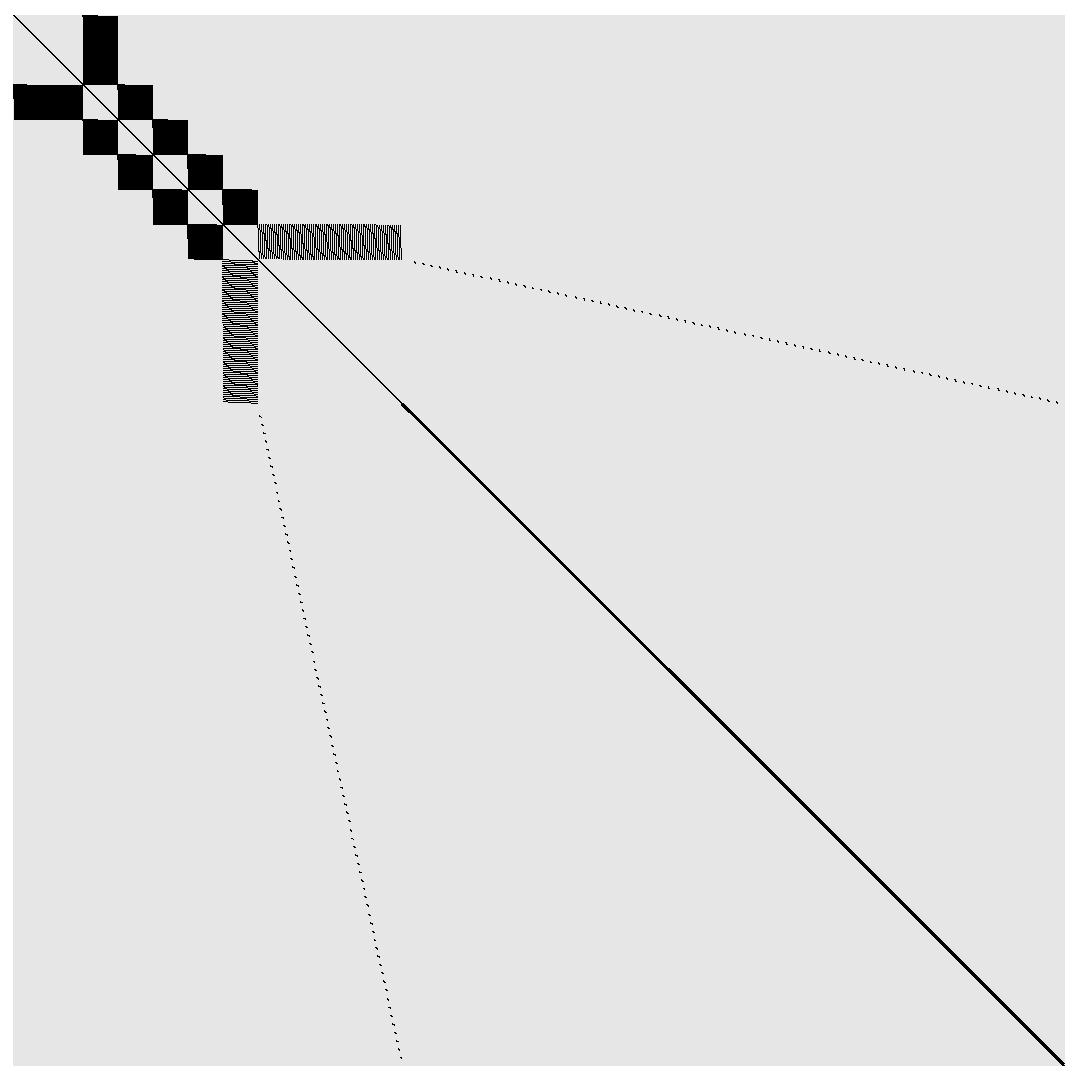}\\
\includegraphics[width=7cm]{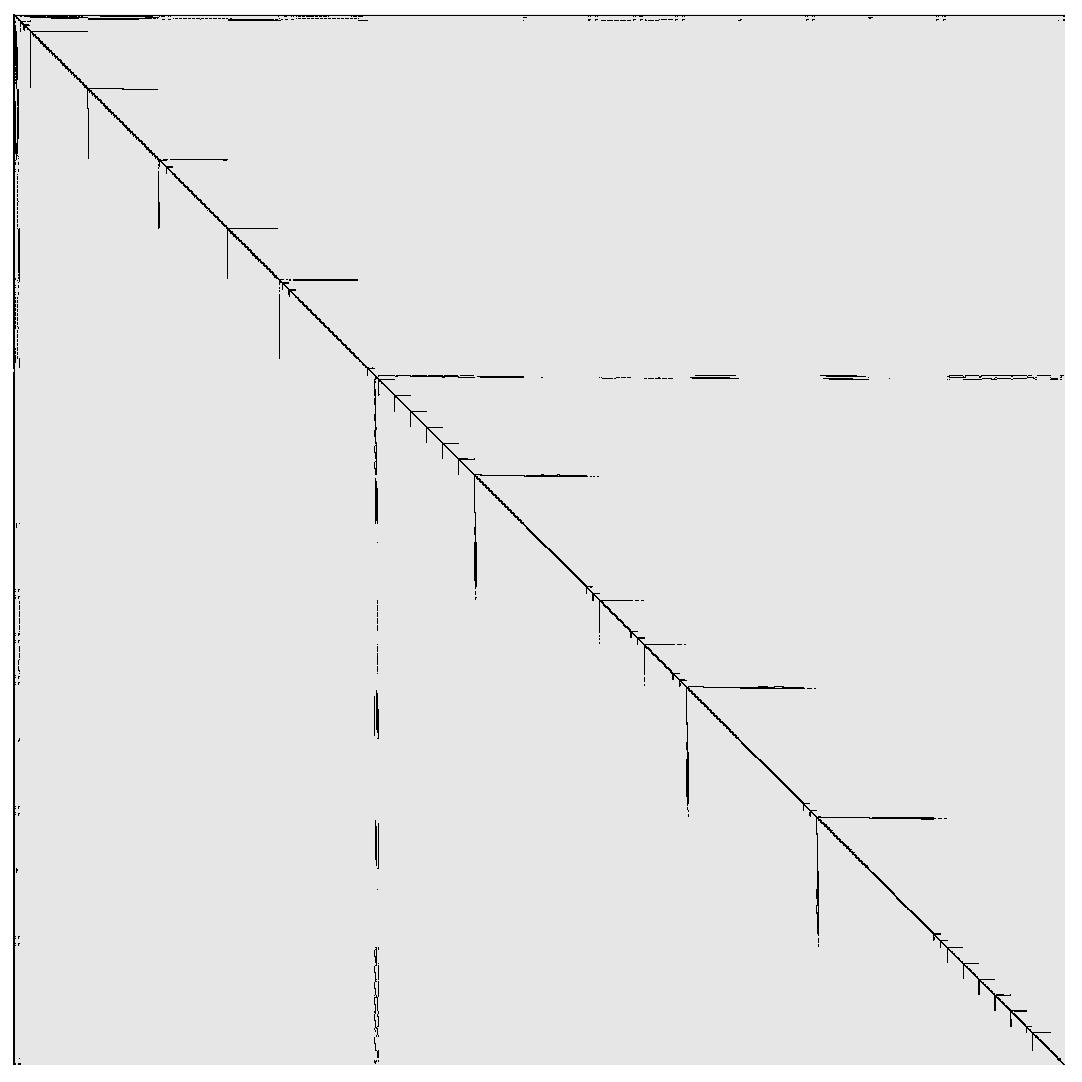}
\includegraphics[width=7cm]{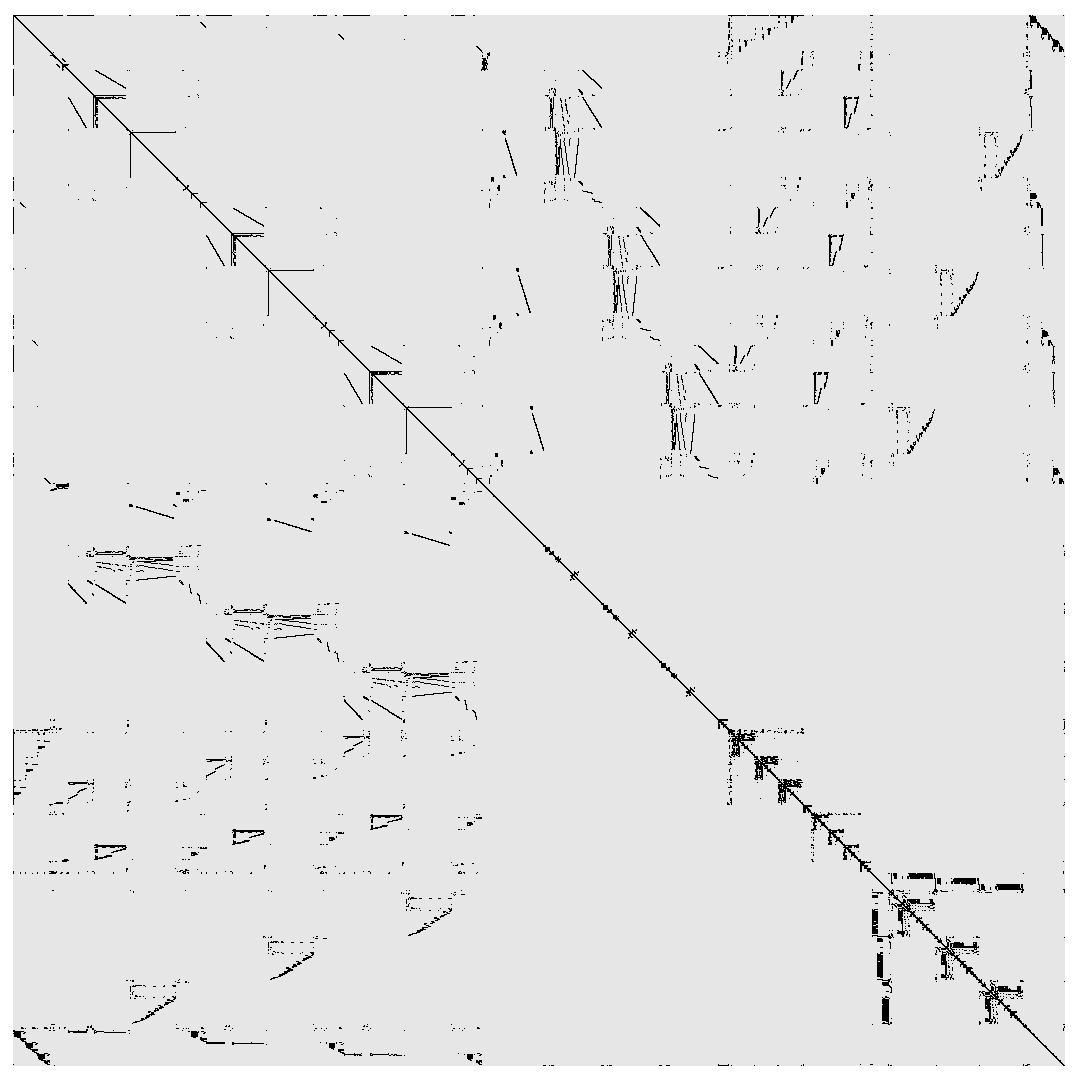}
}
\caption{ Non-zero patterns of matrices: %
  the top-left is {\tt Hohn/fd18}, %
  the top-right is {\tt AMD/G2\_circuit}, %
  the bottom-left is {\tt IBM\_EDA/trans4}, %
  and the bottom-right is {\tt Rajat/Raj1}. %
}
\label{fig:four-matrices}
\end{figure}

%
We show now four matrices from \cite{DavisHu-1994}, 
which produce significant 
differences in performance of the \NEWCSR{} and the Hybrid formats perform.
The matrix names are 
{\tt Hohn/fd18}, {\tt AMD/G2\_circuit}, {\tt IBM\_EDA/trans4}, 
and {\tt Rajat/Raj1},
whose nonzero pattern is shown in Figure \ref{fig:four-matrices}.
Characters of matrices and performances in double precision are 
summarized in Table \ref{tab:four-matices-comparison}.
\NEWCSR{} runs with group size 128 and texture caching on.
\begin{table}[tb]
  \centering
  \begin{tabular}{|l||r|r|r|r||r|r|r||r|}
\hline
Matrix name & \# of & \multicolumn{3}{c||}{\# of nonzeros in row} & \multicolumn{2}{c|}{GFLOPS}& & CPU \\
& rows  & max & mean & min & \NEWCSR{} & Hybrid & ratio & CSR\\
\hline
{\tt Hohn/fd18}       & 16,248  & 6 & 3.860 & 1 &  
4.69  
& 0.95 & 
4.93 & 1.05\\
{\tt AMD/G2\_circuit} & 150,102 & 6 & 4.841 & 2 & 
9.36  
& 2.5 & 
3.74 & 0.60 \\
\hline 
{\tt IBM\_EDA/trans4}   & 116,835 & 114,190 & 6.600 & 1 &
0.019 
& 2.0 & 0.095 & 0.59\\
{\tt Rajat/Raj1}      & 263,743 & 40,468  & 4.938 & 1 & 
0.058 
& 2.2 & 0.026 & 0.50\\
\hline
  \end{tabular}
  \caption{Characters of matrices and performances in double precision with 
the \NEWCSR{} and Hybrid formats, with performance of the CSR format on CPU.}
  \label{tab:four-matices-comparison}
\end{table}
While {\tt Hohn/fd18} and {\tt AMD/G2\_circuit} have small number of 
nonzeros in each row and then \NEWCSR{} performs very well,
{\tt IBM\_EDA/trans4} and {\tt Rajat/Raj1} have large variations in
number of nonzeros and \NEWCSR{} performs very badly.
We call {\tt Hohn/fd18} and {\tt AMD/G2\_circuit} as the first group of the 
four matrices and the rest as the second group.
For more detailed analysis of performance of \NEWCSR{}, we employed
ordering of row index of the matrix.
Employing good ordering of row index, \NEWCSR{} can reduce number of artificial 
nonzeros for alignment of array in each group.
We used the simplest ordering, descending order of number of nonzeros in row,
and AMD ordering (approximate minimum degree ordering) \cite{AmestoyDavisDuff-1996},
which can reduce fill-in during LU factorization. 
Descending ordering is an optimal way to suppress artificial nonzeros but
it may shuffle non-zeros pattern of the matrix.
On the other hand, AMD ordering can reduce range of off-diagonal distribution of
the matrix.
The result is summarized in Table \ref{tab:ordering}. 
The third row of each matrix shows use of texture cache with hit and missed
cases, which was measured by CUDA profiler tool.
Sum of cache hit and missed cases is proportional to number of the nonzeros
of the matrix.
%
We can see ratios of cache miss in the first group are 
even higher than in the second group.
Decreasing ordering can reduce artificial nonzeros drastically
which helps reduction of memory requirement.
AMD ordering shows better use of texture cache than others, but 
suffers from larger artificial nonzeros than descending ordering.
By this comparison of cache miss rate on two groups of matrices, 
we can see ratio of cache hit cases is not the leading term on the performance.
\par
By introducing artificial nonzeros to align size of data in columns per
each group of rows, memory access inside of each multiprocessor is coalesced,
but memory access among multiprocessors is still unaligned 
in the second group matrices due to large variation in numbers of nonzero.
This is the reason why the second group matrices suffer very poor performance
even much worse than common CSR fromat on CPU.
Use of large size of group to achieve aligned access by $30$ multiprocessors, 
such as $128\times 30$, leads unfortunately to very large number of artificial nonzeros.
This is a true weak point of the \NEWCSR{} format.
For matrices enjoying good performance by the \NEWCSR{} format performs well, there is a possibility
of futher improvement of performance 
by decreasing number of artificial nonzeros and
increasing cache utilization by means of ordering of row index.
%
\begin{table}[ht]
  \centering
  \begin{tabular}{|c||r|r|r|}
\hline
&  \multicolumn{3}{|c|}{ordering} \\
 &\mbox{}\hspace{5mm} without \hspace{5mm}\mbox{}& 
\mbox{}\hspace{5mm}descending\hspace{5mm}\mbox{}  & 
\mbox{}\hspace{5mm}AMD\hspace{5mm}\mbox{} \\
\hline
\hline
 \multicolumn{2}{|l}{\tt Hohn/fd18}&\multicolumn{2}{l|}{16,248 rows,\ 63,406 nonzers} \\
\hline
 Artif. zeros & 2.76\% & 0.34\% & 26.38\%\\
                 GFLOPS & 4.690 & 4.845 & 3.900 \\
                 cache hit/miss & 289/1,350 & 324/1,325 & 312/2,907 \\
\hline\hline
 \multicolumn{2}{|l}{\tt AMD/G2\_circuit}&\multicolumn{2}{l|}{ 150,102 rows,\ 726,674 nonzers} \\
\hline
 Artif. zeros & 3.90\% & 0.05\% & 2.03\%\\
                 GFLOPS & 9.364 & 9.210 & 8.044 \\
                 cache hit/miss & 431/12,026 & 622/11,592 & 3,448/11,951\\
\hline\hline 
 \multicolumn{2}{|l}{\tt IBM\_EDA/trans4} & \multicolumn{2}{l|}{116,835 rows,\ 766,396 nonzers} \\
\hline
Artif. zeros & 2,118.1\% & 1,452.8\% & 1,613.3\%\\
                 GFLOPS & 0.0189 & 0.0191  & 0.0188 \\
                 cache hit/miss & 9,213/16,347 & 6,517/12,576 & 11,295/14,727\\
\hline\hline 
 \multicolumn{2}{|l}{\tt Rajat/Raj1} & \multicolumn{2}{l|}{ 263,743 rows,\ 1,302,464 nonzers} \\
\hline
       Artif. zeros & 938.2\% & 189.3\% & 370.6\%\\
                 GFLOPS & 0.0578 & 0.0904 & 0.0808 \\
                 cache hit/miss & 9,229/17,049 & 6,653/12,513 & 10,793/11,707\\
\hline\hline
  \end{tabular}
  \caption{Effect of ordering to number of artificial nonzeros and preformance
of \NEWCSR{}.}
  \label{tab:ordering}
\end{table}
\section{Conclusion}
We proposed \newCSR{} format to store sparse matrix, which can run 
efficiently on the GPU with continuous data access called as coalesced access
in the terminology of the CUDA GPU architecture.
We verified \NEWCSR{} can perform better than the Hybrid format in general by
numerical experiments using 1,600 matrices.
However, on some matrices, \NEWCSR{} performs very poorly due to 
complicated pattern of nonzero elements of matrix,
even though the Hybrid format can perform closed to its average speed.
For enhancement of performance of \NEWCSR{}, good ordering of 
index of the matrix row is necessary, by which usage of texture cache fetching
the right-hand vector is improved. This will be subject of our future research.
\par
The source code of the \NEWCSR{} format is available as a part of the Template Numerical Library (TNL) at 
\href{http://geraldine.fjfi.cvut.cz/~oberhuber/doku-wiki-tnl}{http://geraldine.fjfi.cvut.cz/\textasciitilde  oberhuber/doku-wiki-tnl}.
%
\section*{Acknowledgement}
This work was partially supported by the Jind\v{r}ich Ne\v{c}as Center for Mathematical Modelling, Research center of the Ministry of Education of the Czech Republic LC06052, Research Direction Project of the Ministry of Education of the Czech Republic No. MSM6840770010, and Supercomputing Methods in Mathematical Modelling of Problems in Engineering and Natural Sciences, project of the Student Grant Agency of the Czech Technical University in Prague No. 283 OHK4-009/10 P3913.





\bibliographystyle{elsarticle-num}







\end{document}